\documentclass{aa}
\usepackage{graphicx}
\def\note #1]{{\bf #1]}}
\newcommand{\m}{\,{\rm m}}
\newcommand{\s}{\,{\rm s}}
\newcommand{\Msun}{\hbox{$\rm\thinspace M_{\odot}$}}
\newcommand{\Lsun}{\hbox{$\rm\thinspace L_{\odot}$}}

\newcommand{\teff}{\ensuremath{T_{\rm eff}}}             % T_eff
\newcommand{\kms}{\ensuremath{\,\rm{km\,s}^{-1}}}
\newcommand{\ms}{\ensuremath{\,\rm{m\,s}^{-1}}}
\begin{document}
\title{Detection of Solar-like Oscillations in the G7 Giant Star
$\xi$\,Hya\thanks{Based on observations obtained with the CORALIE spectrograph
on the 1.2-m Swiss Euler telescope at La Silla, Chile.}}

%   \subtitle{I. Observations}

   \author{S. Frandsen
          \inst{1}
          \and
          F. Carrier\inst{2}
          \and
          C. Aerts\inst{3}
          \and
          D. Stello\inst{1}
          \and
          T. Maas\inst{3}
          \and 
          M. Burnet\inst{2}
          \and
          H. Bruntt\inst{1}
          \and
          T.~C. Teixeira\inst{4,1}
          \and\\
          J.~R. de Medeiros\inst{5}
          \and
          F. Bouchy\inst{2}
          \and
          H. Kjeldsen\inst{1,6}
          \and
          F. Pijpers\inst{1,6}
          \and
          J. Christensen-Dalsgaard\inst{1,6}
          }

   \offprints{S. Frandsen}

   \institute{Institut for Fysik og Astronomi, Aarhus Universitet,
              Universitetsparken, DK-8000 Aarhus C, Denmark\\
              \email{srf@ifa.au.dk}
         \and
             Observatoire de Gen\`eve, 51 chemin de Maillettes,
             CH-1290 Sauverny, Switzerland \\
             \email{fabien.carrier@obs.unige.ch}
         \and
              Katholieke Universiteit Leuven, 
Instituut voor Sterrenkunde, Celestijnenlaan 200 B, B - 3001 Leuven, Belgium\\
              \email{conny@ster.kuleuven.ac.be}
\and
Centro de Astrof\'\i{}sica da Universidade do Porto, Portugal
\and
Departamento de F\'{\i}sica, 
Universidade Federal do Rio Grande do Norte, 59072-970,
                  Natal, RN, Brazil
\and
Teoretisk Astrofysik Center, Danmarks Grundforskningsfond
                  }

   \date{Received ; accepted }

   \abstract{We report the firm discovery of solar-like oscillations in a
   giant star. We monitored the star $\xi$~Hya (G7III) continuously during one
   month with the CORALIE spectrograph attached to the 1.2m Swiss Euler
   telescope. The 433 high-precision radial-velocity measurements clearly reveal
   multiple oscillation frequencies in the range 50 -- 130 $\mu$Hz, corresponding
   to periods between 2.0 and 5.5 hours. The amplitudes of
   the strongest modes are 
   slightly smaller than $2\ms$. Current model calculations are compatible with
   the detected modes. \keywords{asteroseismology -- solar type oscillations --
   giant stars}} \maketitle
%
%________________________________________________________________

\section{Introduction}

   Doppler studies with high-precision instruments and reduction algorithms have
   been refined dramatically, mainly in the framework of the search for
   exo\-planets.  These refinements have led to a breakthrough in the
   observations of solar-type oscillations, which have now been found repeatedly
   (Procyon, Martic et al.\ \cite{martic}; $\beta$~Hyi, Bedding et al.\
   \cite{bedding}; $\alpha$~Cen\,A, Bouchy \& Carrier \cite{bouchy3};
   $\delta$~Eri, Carrier et al.\ \cite{carrier}).  The signal-to-noise ratio
   (S/N) in the oscillation frequency spectra is, for each of these cases, so
   good that the resemblance with the solar oscillation spectrum is obvious.

   Observations of solar-like oscillations in the giant star $\alpha\,$UMa have
   been claimed by Buzasi et al.\ (\cite{buzasi}), based upon space photometry
   gathered with the star tracker of the WIRE satellite.  However, the
   interpretation of these reported oscillations frequencies is not
   straightforward. Guenther et al. (\cite{guenther}) find a possible solution
   in terms of a sequence of radial modes with a few missing orders for a star
   of 4.0--4.5 \Msun. The interpretation is not supported by theoretical
   calculations by Dziembowski et al. (\cite{dziembowski}).
   Velocity observations of Arcturus provide evidence for solar-type
   oscillations with periods from 1.7 to 8.3 days and a frequency separation of
   evenly spaced modes of $\Delta \nu \sim 1.2\ \mu$Hz (Merline, 1999).
   WIRE data (Retter et~al. \cite{retter}), however, points to an excess
   power at 4.1~$\mu$Hz and a frequency spacing of $\Delta \nu = 0.8~\mu$Hz.

   In this {\it Letter\/}, we provide clear evidence for the
presence of solar-type oscillations in the giant star $\xi$~Hya ($m_V=3.54$).
This star has a mass close to $M = 3.0 \Msun$, and is thus considerably heavier
than the Sun. Moreover, its luminosity amounts to $L \sim 61 \Lsun$ and its
effective temperature $\teff = 5000$~K, which places the star among the giants.
In the current {\it Letter\/} we present the first results of our study.
Detailed modelling will be presented, when completed, in a subsequent paper.
%in the main journal.

%-----------------------------------------------------------------------------
\section{The radial-velocity measurements}

\begin{figure}
\resizebox{\hsize}{!}{\includegraphics{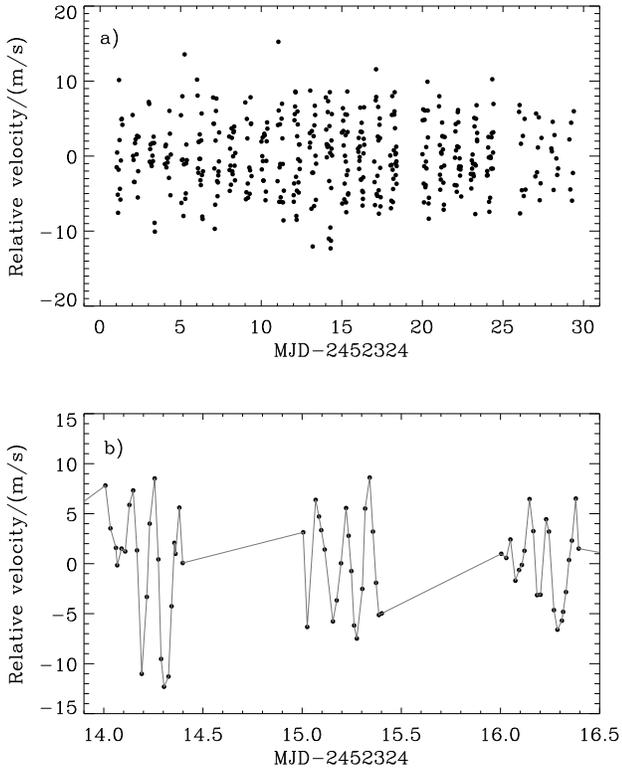}}
\caption{Radial-velocity measurements of $\xi$~Hya. The dispersion
reaches $4.6\m\s^{-1}$. 
{\bf a)} The whole time string; {\bf b)} zoom of a small
part of the time series.
}\label{fig:rv}
\end{figure}

The $\xi$~Hya observations were made during one full month (2002 February 18 -
March 18) with \textsc{Coralie}, the high-resolution fiber-fed echelle
spectrograph (Queloz et al.\ \cite{queloz}) mounted on the 1.2-m Swiss telescope
at La Silla (ESO, Chile). During the stellar exposures, the spectrum of a
thorium lamp carried by a second fiber is simultaneously recorded in order to
monitor the spectrograph's stability and thus to obtain high-precision
velocity measurements.  The spectra were extracted at the telescope, using the
INTER-TACOS (INTERpreter for the Treatment, the Analysis and the COrrelation of
Spectra) software package developed by D.\ Queloz and L.\ Weber at the Geneva
Observatory (Baranne et al.\ \cite{baranne}).  The wavelength coverage of these
spectra is 3875-6820~\AA , recorded on 68 orders. By taking about 2 measurements
every hour, a total of 433~optical spectra was collected. The exposure times
were typically 180\,s and the S/N ratio for all spectra was in the
range of 110--230 at 550\,nm.

By the use of the optimum-weight procedure (Bouchy et al.\ \cite{bouchy2}),
radial velocities are computed for each night relative to the highest S/N
spectrum obtained in the middle of the night. This method requires a Doppler
shift that remains small compared to the line-width (smaller than $100 \ms$)
(Bouchy \& Carrier \cite{bouchy3}).  Since the Earth's motion can introduce a
Doppler shift larger than $800 \ms$ during a whole night, each spectrum is first
corrected for the Earth's motion before deriving the radial velocities. This is
achieved by shifting all spectra with the Earth's velocity at the time of
observation and by subsequent rebinning, so that the spectra all have the same
wavelength values. From each rebinned spectrum a velocity is derived.
The mean radial velocity for each night is then subtracted.
The resulting velocities are shown in
%Fig.~\ref{fig:rv}. 
Fig.\,1.
The rms scatter of the time series is $4.6 \ms$ and is
largely due to the oscillations. The mean error on each measurement
is $2.3 \ms$.

\section{Time-series analysis}

   \begin{figure} \centering \resizebox{\hsize}{!}{\includegraphics{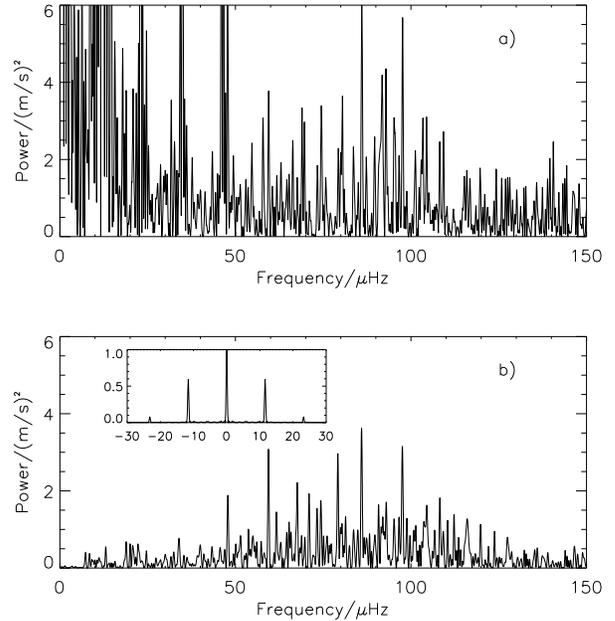}}
   \caption{Power spectrum for $\xi$~Hya.  Modes are clearly present in the
   range 50--130 $\mu$Hz, corresponding to periods of 2.0-5.5 hours. 
   Panel a) is the power spectrum of the time series using one 
   single spectrum as
   the reference. Panel b) is for the time series derived using one
   reference spectrum per night. 
   The insert in Panel b) shows the window function. The
   frequency scale is made similar to the main plot to facilitate
   comparison. } \label{power} \end{figure}

The power spectrum of the 433 data points is presented in Fig.~\ref{power}.  In
order to show that the excess power in the lower panel in the range 50--130
$\mu$Hz is not due to elimination of power at low frequencies by the reduction
procedure, we show a power spectrum in the upper panel where no correction for
drift has been applied. Although more noisy the increase in power between
50~$\mu$Hz and 130~$\mu$Hz is also evident in the upper panel.  A set of
oscillation modes is clearly present, with a power distribution that is
remarkably similar to that seen in the Sun and other stars on or near the main
sequence, although obviously at much lower frequency.  We will show
the spectrum displays the characteristic near-uniform spacing of 
the dominant peaks.

   \begin{figure} \centering
   \resizebox{\hsize}{!}{\includegraphics{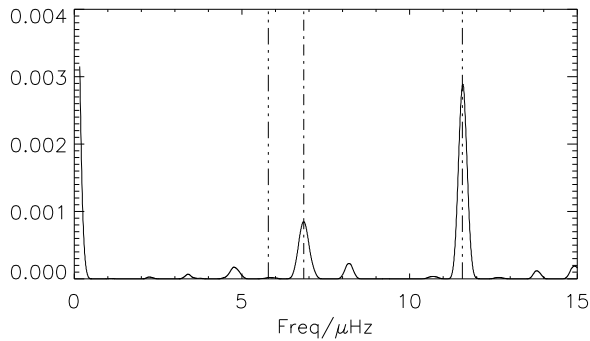}} \caption{Autocorrelation
   of the power spectrum for $\xi$~Hya. All points with amplitude $v < 1.2$~\ms
   have been set to zero. The dash-dot-dot-dot vertical lines mark the position
   of the~11.574 $\mu$Hz daily alias and half of it, and the dash-dot line
   indicates the peak pointing to a frequency spacing of 6.8~$\mu$Hz.}
   \label{auto} \end{figure} 
To characterize this pattern we have calculated the autocorrelation of the power
   spectrum. In order to eliminate the effect of the noise, we have ignored all
   points with amplitudes below a given threshhold (1.2~\ms). In Fig.~\ref{auto}
   the alias at 1\ c/d stands out clearly, but in addition a spacing $\Delta \nu
   \sim 6.8\ \mu$Hz is present in the power spectrum. This is consistent with
   the visual impression of a regular pattern present in the the power spectrum
   (Fig.~\ref{power}).

The expected velocity amplitude for solar-like oscillations scales as $L/M$
according to Kjeldsen \& Bedding (\cite{kjeldsen}). Using the stellar parameters
from Sect.\,4 and a mass of 3.0\,\Msun \,(Sect.\,4) we find $v_{\rm osc,pred} =
4.6 \m \s^{-1}.$ The observed amplitudes in Fig.~\ref{power} are only about one
third of this prediction.  Recent calculations by Houdek \& Gough (2002),
however, indicate that the simple scaling law of Kjeldsen \& Bedding
(\cite{kjeldsen}) does indeed not apply. They predict a velocity amplitude $v =
2.1 \m \s^{-1}$, slightly higher than the ones observed by us for $\xi$~Hya.

The stochastic nature of solar-like oscillations implies that a timestring of
radial velocities cannot be expected to be a set of coherent oscillations and
can therefore not be reproduced perfectly by a sum of sinusoidal terms. As a
starting point it is nevertheless a good assumption to try to fit the
radial-velocity data of $\xi$~Hya by such a set of functions, as the lifetimes
are expected to exceed the length of the observing run (Houdek \& Gough
2002). We have performed an iterative fit using different methods, among which
Period98 (Sperl \cite{sperl}) and a procedure by Frandsen et al.\
(\cite{praesepe}).  An oscillation with a S/N above 4 is detected at 9 frequencies.
When a fit based upon these 9 frequencies is removed from the time series, 5
additional peaks still occur but with a too low S/N to accept them without
additional confirmation (hence we do not list them). After removing the 14
frequencies, only a noise spectrum is left.  The results for the 9 frequencies
are presented in Table~\ref{f_tab}, where the S/N indicated is calculated from
the remaining noise in the amplitude spectrum at the position of each mode.  The
noise is slightly higher in the range of the modes than at high frequencies,
where $\sigma = 0.2\,\ms$.  The frequencies of the modes with amplitudes above
$1.5 \ms$, i.e., of the five highest-amplitude modes, are unambiguous.
Dividing the dataset in two, four out of five modes with S/N $> 5.0$ are
present in each set.  For lower S/N the alias problems lead to a risk that false
detections are made.  Modes with S/N $< 5$ must be considered with some caution.
Confirmation of these frequencies is  needed by additional observations.
\begin{table}
\caption{Oscillation frequencies detected in the radial velocities of $\xi$~Hya
ordered by amplitude. Formal errors of the fit are given in parenthesis as
errors of the last digits. The S/N is calculated by dividing the amplitude by
the noise in a 10 $\mu$Hz bin in the residual spectrum centered on the frequency
of the mode. Frequencies in italic indicate modes where the frequency
detection methods disagree on the
correct alias. The last column gives the difference between the
observed frequency and the frequency derived from Eq.~(1).}\label{f_tab}
\begin{tabular}{rrr|r|r|r}
\hline
ID & \multicolumn{2}{c|}{Frequency} & Amplitude & S/N & $\Delta \nu$\\
 &  c/d & $\mu$Hz & \ms & & $\mu$Hz \\
\hline
F1 & 5.1344(26) & 59.43 & 1.85(23) & 6.6 & 0.77 \\
F2 & 6.8366(27) & 79.13 & 1.84(23) & 5.8 & -0.86 \\
F3 & 7.4265(29) & 85.96 & 1.76(23) & 5.3 & -1.14 \\
F4 & 8.2318(32) & 95.28 & 1.65(23) & 5.1 & 1.07 \\
F5 & 9.3507(33) & 108.22 & 1.59(23) & 6.0 & -0.21 \\
F6 & {\it 8.7399(36)} & 101.16 & 1.36(22) & 4.5 & -0.16 \\
F7 & 10.0287(43) & 116.07 & 1.25(23) & 5.0 & 0.53 \\
F8 & {\it 9.0831(44)} & 105.13 & 1.24(24) & 4.3 \\
F9 & {\it 8.5339(40)} & 98.77 & 1.23(23) & 4.1 \\
% 10 & 5.6426(58) & 65.31 & 1.15(25) & 3.8 \\
% 11 & 5.6813(61) & 65.76 & 1.15(25) &  \\
% 12 & 6.3385(44) & 73.36 & 1.10(23) &  \\
% 13 & 7.9529(45) & 92.05 & 1.08(23) &  \\
% 14 & 11.6868(56) & 135.26 & 0.84(23) & \\
\hline
\end{tabular}
\end{table}
What is stated above has been verified by the analysis of several sets 
of simulated data assuming a variety
of lifetimes in order to check the validity of the identified modes. The
details of such simulations will be published in a subsequent paper.

The first seven modes can be ordered in a sequence of modes, which 
fits the straight line 
\begin{equation}
\nu(n) = 1.78 + 7.11 (\pm 0.14) ~n
\end{equation}
where $n$ is an integer (the order of the mode). Some $n$ values are
missing. 
The maximum deviation from the line is 1.14~$\mu$Hz (Table 1). 
The regularity seen is similar to the results reported for Arcturus (Merline, 1999)
and $\alpha$~UMa (Guenther et al.,~\cite{guenther}).
The first seven modes could all be radial modes, although we
cannot rule out the possibility of alternating degree
$\ell = 0$ and $\ell = 1$ modes. This, however, is not consistent
with the model presented in Sect. 4. 

From the present data, we cannot firmly exclude that frequency F8 in Table 1
corresponds to the same mode as F7. The two modes are resolved, but if the
damping time is short, they might be different realizations of the same mode.

%_____________________________________________________________
\section{First interpretation}

   \begin{figure}
   \centering
  \resizebox{\hsize}{!}{\includegraphics{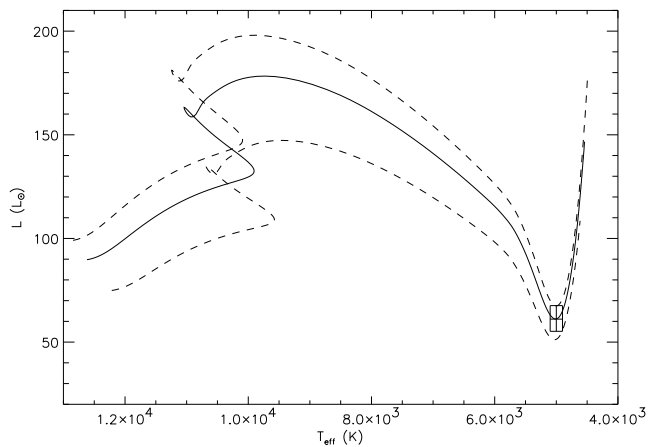}}
      \caption{Hertzsprung-Russell diagram with evolutionary tracks for $Z =
      0.019$ for masses $3.07 M_\odot$ (solid line), $2.93$ and $3.15 M_\odot$
      (lower and upper dashed lines, respectively). The location of $\xi$~Hya in
      the diagram is given by the error box.  } \label{figure2} \end{figure}

In order to study the nature of the oscillations detected in $\xi$~Hya, it is
necessary to compare the observed frequency spectrum with model predictions,
taking into account the constraints on the three observational stellar
parameters $T_{\rm eff}$, $L$ and $Z$. We have redetermined the atmospheric
parameters and use $T_{\rm eff} = 5000\pm100\ $K, $L = 61.1\pm6.2$\ \Lsun,
$\log g = 2.85$, [M/H] = $0.06\pm0.07$ ($Z = 0.019\pm0.006$) and 
$v \sin i = 1.8\pm1.0\ $\kms. Details of how these values were obtained 
will be reported in a subsequent paper.

Using the evolution code of
Christensen-Dalsgaard (1982), evolutionary tracks were produced, spanning the
error box defined by the uncertainties of $T_{\rm eff}$, $L$ and $Z$.
The model tracks were computed using the EFF equation of state 
(Eggleton et al.\ 1973), OPAL opacities (Iglesias, Rogers \& Wilson 1992),
Bahcall \& Pinsonneault (1992) nuclear cross sections, and the mixing-length
formalism (MLT) for convection.

The evolutionary track passing through the observed ($T_{\rm eff}$, $L$)
corresponds to a mass of $3.07 M_\odot$ for $Z = 0.019$
(Fig.~\ref{figure2}). Oscillation frequencies were calculated for the model in
that track closest to the observed location of $\xi$~Hya in the HR~diagram. The
average separation between radial modes in the range 50--100\,$\mu$Hz is $\Delta
\nu_{\rm theo} = 7.2\,\mu$Hz in agreement with the observational value.
The model frequencies fit a linear relation for orders $5 < n < 17$ or
$50 < \nu < 125~\mu$Hz given by $\nu = 7.15 + 7.14~n$
with absolute values only 1--2~$\mu$Hz from the observed frequencies.
The maximum deviation of the model frequencies from the line is $\sim 0.5~\mu$Hz.

 % resulting frequency spectrum is shown in Fig.~\ref{figure3}. Despite the
 % complexity of the predicted model spectrum, a large separation of
 % $\Delta \nu \sim 7.2\,
 % \mu {\rm Hz}$ is clearly identifiable in the region of interest, which is in
 % very good agreement with the observed value
 % quoted in Sect. 4.
 
Radial modes are expected to dominate the spectrum for giant stars (cf.\
Dziembowski et al.\ \cite{dziembowski}, Fig.\ 2).  Further analysis of the spectrum
is beyond the scope of the current discovery paper and will be done
in a forthcoming paper, dealing in detail with the issue of modelling and
considering also the possibility that $\xi$~Hya could be a core
helium burning star with a smaller mass.

%    \begin{figure}
%    \centering
%   \resizebox{\hsize}{!}{\includegraphics{xiHya-fig3.ps}}
%       \caption{Predicted frequency spectrum in the range $50 - 120 \mu
%    {\rm Hz}$ for a $3.07 M_\odot$ model that matches the observed parameters
%    of $\xi$~Hya. Solid lines correspond to $l = 0$, dotted lines to $l
%    = 1$, dashed lines to $l = 2$.
%    %and dot-dashed lines to $l = 3$.
%    \note [Drop $l = 3$. Remove scaling with 0.7 for $l = 2$, or remark on
%    it!]
%    Amplitudes, given in arbitrary units, were normalized to the
%    corresponding interpolated value for the radial modes ($l = 0$) for
%    any given frequency, assuming fixed total energy in the mode, as
%    done by Christensen-Dalsgaard et al. (1995).  
%               }
%          \label{figure3}
%    \end{figure}
%

\section{Conclusions}

The main conclusions of this study are as follows.  Solar-like oscillations have
been firmly discovered in the bright G7III star $\xi$~Hya. The amplitudes of the
strongest modes are somewhat below $2\ms$.  The observed frequency distribution
of the modes detected (Table 1) is in agreement with theoretically calculated
frequencies both in terms of the spacing and the absolute values.  The modes
with the largest amplitudes can be well matched with radial modes that have an
almost equidistant separation around 7.1~$\mu$Hz.

A most important and exciting result of our study is the confirmation of the
possibility, suggested by the results reported on $\alpha$~UMa and Arcturus, to
observe solar-like oscillations in stars on the red giant branch. This opens the
red part of the HR\,diagram for detailed seismic studies. The latter require an
accuracy within the range of current and future ground-based instruments.  Such
future studies will only be successful if an extremely high stability of the
instrument is achieved and if one performs multisite observing campaigns
covering several months in order to resolve the frequency spectrum of the
oscillations and to eliminate the aliasing problems.

\begin{acknowledgements}

CA and TM acknowledge the Fund for Scientific Research of Flanders under project
G.0178.02 for its financial support of the Leuven contribution to the CORALIE
observations of $\xi$ Hya and of the PhD position of TM.  Part of this work was
supported financially by the Swiss National Science Foundation.  Support was
received as well from the Danish National Science Foundation through the
establishment of the Theoretical Astrophysics Center, from Aarhus University and
from the Danish Natural Science Research Council.  TCT is supported by research
grant SFRH/BPD/3545/2000 of the {\it Funda\c c\~ao para a Ci\^encia e a
Tecnologia}, Portugal.

\end{acknowledgements}

\end{document}